\documentclass[10pt,aps,prd,floatfix,notitlepage,showpacs]{revtex4-1}
\usepackage{amsmath,amssymb}
\usepackage{graphicx}
\usepackage{tikz}
\usetikzlibrary{matrix}
\usepackage{epstopdf}

\begin{document}

\title{Conformal transformations in modified teleparallel theories of gravity revisited}

\author{Matthew Wright}
\email{matthew.wright.13@ucl.ac.uk}

\affiliation{Department of Mathematics, University College London, Gower Street, London, WC1E 6BT, United Kingdom}

\date{\today}

\begin{abstract}
It is well known that one cannot apply a conformal transformation to $f(T)$ gravity to obtain a minimally coupled scalar field model, and thus no Einstein frame exists for $f(T)$ gravity. Furthermore nonminimally coupled ``teleparallel dark energy models'' are not conformally equivalent to $f(T)$ gravity. However, it can be shown that $f(T)$ gravity is conformally equivalent to a teleparallel phantom scalar field model with a nonminimal coupling to a boundary term only. In this work, we extend this analysis by considering a recently studied extended class of models, known as $f(T,B)$ gravity, where $B$ is a boundary term related to the divergence of a contraction of the torsion tensor. We find that nonminimally coupled ``teleparallel dark energy models" are conformally equivalent to either an $f(T,B)$ or $f(B)$ gravity model. Finally conditions on the functional form of $f(T,B)$ gravity are derived to allow it to be transformed to particular nonminimally coupled scalar field models.
\end{abstract}

\maketitle

\section{Introduction}

Theories of gravitation are dynamically equivalent if one can map one theory to another by applying a conformal transformation to the metric. Conformal symmetry is a fundamental symmetry of spacetime and generalises the concept of scale invariance. Within the curvature formulation of gravity, this has been used to show that a well studied modification of gravity, $f(R)$ gravity, is dynamically equivalent to Einstein gravity with an additional matter field, taking the form of a canonical scalar field. This conformal transformation can be chosen so that the scalar field is minimally coupled to the gravitational sector, so that there is no $A(\phi)R$ terms within the action, where $A(\phi)$ is a generic function of the scalar field $\phi$. This equivalence is often exploited to great power, as one can choose which frame to perform calculations in and derive physical quantities based on which frame it is simpler to do so. It is also known that there is a conformal relationship between phantom scalar field theories, where the kinetic energy of the scalar field has the incorrect sign, and complex $f(R)$ gravity~\cite{Briscese:2006xu}. 

Teleparallel gravity is an alternative formulation of gravity that is based on the torsion tensor. In general relativity the connection, the Levi-Civita connection, possesses curvature but is absent of torsion. However, it can be shown that if one chooses an alternative connection, known as the Weitzenb\"ock connection, which possesses zero curvature, but non-trivial torsion,  one can build a theory which is dynamically equivalent to general relativity. This theory is known as the teleparallel equivalent of general relativity (TEGR), and was first formulated by Einstein in the 1920s. Despite TEGR's equivalence to general relativity, it has a very different physical interpretation. In general relativity, the gravitational interaction is described by curvature, which is used to geometrize spacetime. However in teleparallel gravity, the gravitational interaction is described by torsion, but this interaction in not geometrized, rather it behaves as a force. 

In recent years there has been an increasing amount of work studying modifications within the teleparallel framework of gravity. This has been motivated by their ability to provide a possible explanation for the late time accelerated expansion of the universe. $f(T)$ gravity, where $T$ is the torsion scalar, is a well studied modification of gravity within this class~\cite{Ferraro:2006jd,Bengochea:2008gz}, with such models providing different dynamics to $f(R)$ gravity. One advantage of $f(T)$ gravity as opposed to $f(R)$ gravity is that the field equations remains second order, yet the price you pay for this is lack of local Lorentz invariance in the theory~\cite{Li:2010cg} (although see the following recent works for a new approach to this problem by choosing a non-vanishing spin connection~\cite{Krssak:2015lba,Krssak:2015oua}). Another teleparallel modification which has been considered in the last few years has been to consider a scalar field nonminimally coupled to torsion, this is sometimes known as {\it teleparallel dark energy}, first introduced in~\cite{Geng:2011aj}. This model has been shown to exhibit interesting late time cosmological phenomenology, with accelerating attractor solutions and the possibility of a dynamical crossing of the phantom barrier. 

Recently the consideration of a boundary term $B$, which is related to the divergence of the torsion vector
\begin{align}
B=2\nabla_\mu T^\mu
\end{align}
has led to a study of a broader class of modifications~\cite{Bahamonde:2015zma,Bahamonde:2015hza}. This is inspired by the relationship between the torsion scalar $T$ of teleparallel gravity and the Ricci scalar of general relativity, with the boundary term being the difference between them. $f(T,B)$ gravity was studied in~\cite{Bahamonde:2015zma}, where the relationship between $f(R)$ gravity and $f(T)$ gravity was examined. A further consideration of this term looked at the effect of a nonminimal coupling of a scalar field to this boundary term~\cite{Bahamonde:2015hza} in the context of late time accelerating solutions, and some interesting cosmological phenomenology was derived. 

The purpose of this paper is to consider the conformal equivalence between some of these different modifications in the teleparallel framework. $f(T)$ gravity, unlike $f(R)$ gravity is known not to be equivalent to Einstein gravity with a minimally coupled scalar field~\cite{Yang:2010ji}. Instead an additional torsion scalar field coupling is found to be present. In this work we take another look at this, and show that this additional coupling can take the interpretation of a nonminimal coupling between the scalar field and the boundary term $B$, similar to the model studied in~\cite{Bahamonde:2015hza}. However, as observed in~\cite{Li:2010cg}, the kinetic energy of this nonminimally coupled scalar field has the incorrect sign, which can lead to instabilities at the level of perturbations. 

Furthermore, we look at the conformal behaviour of coupled $A(\phi)T$ theories and show that they can be mapped to a particular class of $f(T,B)$ theories of gravity. We then start with an $f(T,B)$ gravity, and examine conditions on the functional form of $f$ required to map to particular types of minimial and nonminimally coupled scalar field theories. We also look at a particular subset of these models, which we call $f(B)$ gravity, where $f(T,B)=-T+f(B)$, and are able to show that these are conformally equivalent to a teleparallel dark energy theory, where only a nonminimal coupling between a scalar field and the torsion scalar $T$ is present. 

This paper is organised as follows. In Section II we introduce the teleparallel formulation of gravity and outline the conventions and notation used throughout. In Section III we review conformal transformations, looking at how $f(R)$ gravity can be transformed to an Einstein frame, and look at how torsion based tensors transform under conformal transformations. In Section IV we look at how $f(T)$ gravity can be transformed to a scalar frame, while in 
Section V we look at how teleparallel dark energy can be mapped to an $f(T,B)$ theory. In Section VI we explore whether $f(T,B)$ gravity can be transformed to a scalar frame, and in Section VII we consider $f(B)$ gravity. Finally in Section VIII we summarise our results, with the main relationships found displayed in Figure~\ref{pic}.

\section{Teleparallel gravity}

In this section we introduce teleparallel gravity. The fundamental objects in teleparallel gravity are the tetrads, which is a basis of orthonormal vectors on the tangent space, and we denote them by $\mathbf{e}_a(x^\mu)$. Here the Latin indices are coordinates on the tangent space, whereas Greek indices correspond to spacetime coordinates and both takes values $a, \mu=0,1,2,3$. The orthonormality condition means that
\begin{align}
\mathbf{e}_a \cdot \mathbf{e}_b= \eta_{ab},
\end{align}
where $\eta_{ab}$ is the Minkowski metric with signature $(+,-,-,-)$, so that
\begin{align}
\eta_{ab}=\textrm{diag} (1,-1,-1,-1).
\end{align} 
The tetrad in component form in terms of the coordinate basis of the manifold is denoted as
\begin{align}
\mathbf{e}_a=e^{\mu}_a \partial_{\mu}.
\end{align}
The {\it inverse tetrad} or {\it cotetrad} $e_{\mu}^{a}$ is defined such that both
\begin{align}
  e_{m}^{\mu} e_{\mu}^{n} &= \delta^{n}_{m} \,,
  \label{deltanm} \\
  e_{m}^{\nu} e_{\mu}^{m} &= \delta^{\nu}_{\mu} \,.
  \label{deltamunu}
\end{align}

The physical metric $g_{\mu\nu}$ on the manifold can be expressed in terms of the tetrad and the Minkowski metric as so
\begin{align}
  g_{\mu\nu} &= e^{a}_{\mu} e^{b}_{\nu} \eta_{ab} \,,
\end{align}
and likewise the inverse metric can also be expressed in terms of the cotetrads
\begin{align}
  g^{\mu\nu} &= e^{\mu}_{a} e^{\nu}_{b} \eta^{ab} \,,
\end{align} 
The quantity $e$ is defined to be the determinant of the tetrad $e^a_\mu$, which can easily be seen to be equivalent to the volume element of the metric, so that $e=\sqrt{-g}$.

Equipped on our manifold we have a connection which we denote by $\Gamma^\lambda_{\nu\mu}$. We define the {\it torsion tensor} to be the antisymmetric part of the connection
\begin{align}
T^\lambda{}_{\mu\nu}:=\Gamma^\lambda_{\nu\mu}-\Gamma^\lambda_{\mu\nu},
\end{align}
with the torsion vector $T_\mu$ being defined as the following contraction of the torsion tensor
\begin{align}
T_\mu=T^\lambda{}_{\lambda\mu}.
\end{align}
Then the {\it contorsion tensor} $K$ is defined as the difference between the connection and the standard Levi-Civita connection, which we denote by $\bar{\Gamma}^\lambda_{\mu\nu}$, so that
\begin{align}
K^\lambda{}_{\mu\nu}:= \Gamma^\lambda_{\mu\nu}-\bar{\Gamma}^\lambda_{\mu\nu}.
\end{align}
The contorsion tensor can also be found to be equivalent to the following combinations of the torsion tensor
\begin{align}
K^\lambda{}_{\mu\nu}=\frac{1}{2}(T_\mu{}^\lambda{}_\nu+T_\nu{}^\lambda{}_\mu-T^\lambda{}_{\mu\nu}). \label{contortion}
\end{align}

We also introduce the tensor $S$, sometimes known as the {\it superpotential} which is the following combination of torsion and contorsion
\begin{align}
S^{\rho\mu\nu}:=\frac{1}{2}\left(K^{\mu\nu\rho}-g^{\rho\nu}T^{\sigma\mu}{}_\sigma+g^{\rho\mu}T^{\sigma\nu}{}_{\sigma}\right). \label{superpotential}
\end{align}
Associated with the superpotential is the following invariant, sometimes known as the torsion scalar $T$
\begin{align}
T= S^{\rho\mu\nu}T_{\rho\mu\nu},
\end{align}
which forms the basis for the Lagrangian of teleparallel gravity.

In the Teleparallel Equivalent of General Relativity one works with the Weitzenb\"ock connection. This particular choice of connection is equivalent to choosing a vanishing spin connection, and is given explicitly in terms of the tetrad fields as
\begin{align}
\Gamma^{\lambda}_{\mu\nu}=e_b^\lambda \partial_\nu e^b_\mu= -e^b_\mu \partial_\nu e^\lambda_b. \label{Weitzen}
\end{align}
Calculating the Riemann curvature tensor associated with this Weitzenb\"ock connection, it is found that it vanishes identically,
\begin{align}
R^\rho{}_{\mu\lambda\nu}=0,
\end{align}
and so the spacetime is globally flat. However in general the connection has a non-vanishing torsion. 

The action of the teleparallel equivalent of general relativity (TEGR) is given by the following integral of the torsion scalar
\begin{align}
S= -\frac{ 1}{16\pi G}\int  T e\, d^4x \, . \label{TEGR}
\end{align}
Varying this action with respect to the tetrad generates the field equations, which are found to be equivalent to Einstein's field equations. At the level of the action, the dynamical equivalence of this action to general relativity can be easily seen by deriving the following relationship between the Ricci scalar $R$ of the Levi-Civita connection and the torsion scalar
\begin{align}
R=-T+B, \label{RTB}
\end{align}
where $B$ is the following boundary term
\begin{align}
B=2\nabla_\mu T^\mu=\frac{2}{e} \partial_\mu (eT^\mu).
\end{align}
Any linear dependence on $B$ in the action will  not effect the dynamics of the system. This relationship shows that the action~(\ref{TEGR}) is dynamically equivalent to the standard Einstein Hilbert action, since they only differ by a total derivative.

There are various interesting ways that teleparallel gravity can be modified. Of interest to this paper are $f(T)$ gravity~\cite{Ferraro:2006jd}, which has the following action
\begin{align}
S_{f(T)}= -\frac{ 1}{16\pi G}\int   f(T) e\, d^4x \, , \label{fT}
\end{align}
and $f(T,B)$ gravity~\cite{Bahamonde:2015zma}, which has the following action (note the difference in the signs between the two actions, which is just a matter of convention)
\begin{align}
S_{f(T,B)}= \frac{ 1}{16\pi G}\int   f(T,B) e\, d^4x \,. \label{fTB}
\end{align}
$f(T,B)$ gravity is a general framework which encompasses both $f(T)$ gravity and $f(R)$ gravity in suitable limits. In particular when the function form of $f(T,B)=f(-T+B)$ we recover $f(R)$ gravity due to the relationship~(\ref{RTB}).

Also of interest to this paper are the equivalent of the scalar-tensor theories in the teleparallel framework. If one considers a canonical scalar field minimally coupled to the gravitational sector, one recovers standard general relativity with a scalar field. However, if one considers a nonminimal coupling, a different theory is obtained to the corresponding curvature based formulation. For example, the action
\begin{align}
S= \frac{ 1}{16\pi G}\int  \left[ -A(\phi)T-\frac{1}{2}\partial_\mu \phi \partial^\mu \phi-V(\phi)+\mathcal{L}_m \right] e d^4x,
\end{align}
where a scalar field is coupled nonminimally to the torsion scalar. This action has recently been of great interest due to its potential late time cosmological applications~\cite{Xu:2012jf,Wei:2011yr,Otalora:2013dsa,Otalora:2013tba,Skugoreva:2014ena}. 

One could consider other forms of coupling  between a scalar field and torsion tensors. Recently, the consideration of coupling a scalar field to the boundary term has been considered in the context of late time cosmology~\cite{Bahamonde:2015hza,Zubair:2016uhx},  like so
\begin{align}
\frac{ 1}{16\pi G}\int  \left[ -A(\phi)T-C(\phi)B-\frac{1}{2}\partial_\mu \phi \partial^\mu \phi-V(\phi)+\mathcal{L}_m \right] e d^4x,
\end{align}
where there is a nominimal coupling to the boundary term $C(\phi)B$. However, we should note that such a coupling is equivalent, via an integration by parts, to a coupling term of the form
\begin{align}
D(\phi)T^\mu\partial_\mu \phi,
\end{align}
where $D(\phi)=C'(\phi)$. Such a term has also been considered several times, going back to~\cite{Saez}, who considered a Brans-Dicke type coupling in the context of M\o{}ller's tetradic theory~\cite{Moller}. More recently such a coupling has also been considered in~\cite{Bamba,Otalora:2014aoa}.

\section{Conformal transformations}
In this section we introduce the concept of a conformal transformation, and review the relationship between the Einstein and the Jordan frames of $f(R)$ gravity. A conformal transformation $g\rightarrow \hat{g}$ is simply a rescaling of the metric where we  multiply the metric by a scalar field $\Omega$ which is dependent on the spacetime coordinates $x^\mu$, called the conformal factor,  as so
\begin{align}
\hat{g}_{\mu\nu}=\Omega^2(x) g_{\mu\nu}, \quad \hat{g}^{\mu\nu}=\Omega^{-2}(x) g^{\mu\nu}. \label{conformal}
\end{align}
$\Omega$ is required to be real so that the conformal transformation is positive definite. Such a transformation preserves the causal structure of the spacetime, and so theories related to each other via a conformal transformation are dynamically equivalent. 

In the case of modified curvature theories, based on the Ricci scalar, it is well known that $f(R)$ gravity is equivalent to a scalar tensor theory. Let us review how this works. We start with a generic $f(R)$ theory of gravity, which has the following action
\begin{align}
  S_{f(R)} = \frac{ 1}{16\pi G}  \int  f(R) \sqrt{-g}\, d^4x \,. \label{f(R)}
\end{align}
We now introduce two new auxiliary field $\phi$ and $\chi$, so that we can write the equivalent action 
\begin{align}
S= \frac{ 1}{16\pi G}  \int \left[\chi(R-\phi)+f(\phi) \right]\sqrt{-g}\, d^4x \,. \label{f(R)aux}
\end{align}
Varying this action with respect to $\chi$ sets the field $\phi=R$ and we recover~(\ref{f(R)}). Now we can eliminate $\chi$ from this action by varying with respect to $\phi$ and to find that $\chi=f'(\phi)$. We therefore arrive at the following equivalent action
\begin{align}
S= \frac{ 1}{16\pi G}  \int \left[f'(\phi)(R-\phi)+f(\phi) \right]\sqrt{-g}\, d^4x \, , \label{f(R)aux2}
\end{align}
which is an action of a scalar tensor theory.

Now we will apply a conformal transformation to the metric to transform to a minimally coupled scalar frame. In this case we will choose the particular conformal factor
\begin{align}
\hat{g}_{\mu\nu}= f'(\phi) g_{\mu\nu}.
\end{align}
Such a transformation modifies the Ricci scalar $R\rightarrow \hat{R}$. The transformation law for the connection, Ricci scalar and other objects can be found for example in~\cite{Faraoni:1998qx}.

After performing such a transformation, and defining a new scalar field $\sigma$ as the following function of $\phi$
\begin{align}
\sigma= \sqrt{3}\ln f'(\phi), \label{Asigma}
\end{align}
the action~(\ref{f(R)aux2}) takes the form
\begin{align}
S= \int \frac{ 1}{16\pi G}\left[ \hat{R}-\frac{1}{2} \hat{g}^{\rho\nu}\partial_{\rho}\sigma \partial_{\nu}\sigma-V(\sigma) \right]   \sqrt{-\hat{g}}\, d^4x \,.
\end{align}
with the potential $V(\sigma)$ given by
\begin{align}
V(\sigma)=\frac{\phi}{f'(\phi)}-\frac{f(\phi)}{f'(\phi)^2}
\end{align}
which can be rewritten in terms of $\sigma$ by inverting the relation~(\ref{Asigma}). This frame is referred to as the {\it Einstein frame}, where the theory takes the form of a scalar tensor theory in which the gravitational sector and the scalar field are minimally coupled and the action of the scalar field is of the canonical form.

Now let us look at conformal transformations in the teleparallel framework. Under the conformal transformation~(\ref{conformal}), it is easy to see that the tetrad and the inverse tetrad must transform as~\cite{Obukhov:1982zn,Yang:2010ji,Bamba}
\begin{align}
\hat{e}^a_\mu =\Omega(x) e^a_{\mu}, \quad \hat{e}_a^\mu =\Omega^{-1}(x) e^{\mu}_a.
\end{align}
The volume element $e$ transforms as expected under a rescaling of the metric
\begin{align}
\hat{e}=\Omega^4 e.
\end{align}

Now we need to examine how the various other torsion tensors and scalars transform. In a general metric affine framework, the connection and metric are independent quantities, and so the connection can in principle vary arbitrarily under a conformal transformation, hence the torsion tensor transformation law is undetermined.  In a Riemann-Cartan spacetime an additional geometry is imposed which determines the transformation law of the connection and torsion tensor under a conformal rescaling~\cite{Obukhov:1982zn}. Teleparallel gravity is a special case of the Riemann-Cartan space, thus under such a conformal transformation, it is easily calculated that the torsion tensor transforms as~\cite{Obukhov:1982zn,Yang:2010ji}
\begin{align}
\hat{T}^\rho{}_{\mu\nu}=T^\rho{}_{\mu\nu}+\Omega^{-1}(\delta^\rho_\nu \partial_\mu \Omega-\delta^\rho_\mu \partial_\nu \Omega), \label{ttransform}
\end{align}
which can be derived by writing out the torsion tensor in terms of the tetrad.  Using the relation~(\ref{contortion}), this implies that the contortion tensor transforms as
\begin{align}
\hat{K}^{\mu\nu}{}_\rho=\Omega^{-2}K^{\mu\nu}{}_\rho  + \Omega^{-3}(\delta^\nu_\rho \partial^\mu \Omega-\delta^\mu_\rho \partial^\nu \Omega).
\end{align}
Together these imply that the superpotential transforms as
\begin{align}
\hat{S}_\rho {}^{\mu\nu}=\Omega^{-2} S_\rho{}^{\mu\nu}+\Omega^{-3}(\delta^\mu_\rho \partial^\nu \Omega-\delta^\nu_\rho \partial^\mu \Omega).
\end{align}

Contracting these transformations, this allows one to calculate how the torsion scalar transforms
\begin{align}
\hat{T}=\Omega^{-2} T+4\Omega^{-3} g^{\mu\nu}\partial_\nu \Omega T^{\rho}{}_{\rho\mu}-6\Omega^{-4}g^{\mu\nu}\partial_\mu \Omega \partial_\nu \Omega,
\end{align}
from which the inverse transformation can be derived
\begin{align}
T=\Omega^2 \hat{T}-4\Omega \hat{g}^{\mu\nu}\partial_\mu \Omega \hat{T}_\nu-6 \hat{g}^{\mu\nu}\partial_\mu \Omega \partial_\nu \Omega.
\end{align}
where we note that partial derivatives remain unchanged under conformal transformations, so that $\partial_\mu=\hat{\partial_\mu}$. Finally for the purposes of this work, we need to investigate how the boundary term $B$ changes under a conformal transformation. We find
\begin{align}
B=\frac{2}{e} \partial_\mu (eT^\mu)=\frac{2 \Omega^4}{\hat{e}} \partial_\mu (\hat{e}(\Omega^{-2}\hat{T}^\mu+3\Omega^{-3} \hat{\partial}^\mu \Omega)),
\end{align}
where, by contracting~(\ref{ttransform}) we have used that the vector $T^\mu$ transforms as
\begin{align}
T^\mu=\Omega^2 \hat{T}^\mu+3\Omega \hat{\partial}^\mu \Omega.
\end{align}
Expanding out the partial derivative, this means that $B$ transforms as follows
\begin{align}
B&= \Omega^2 \hat{B}-4\Omega \hat{T^\mu}\hat{\partial}_\mu \Omega-18\hat{\partial}^{\mu}\Omega \hat{\partial}_\mu \Omega+\frac{6}{\hat{e}}\Omega \hat{\partial}_\mu (\hat{e}\hat{g}^{\mu\nu}\hat{\partial}_\nu\Omega)\nonumber
\\
&=\Omega^2 \hat{B}-4\Omega \hat{T^\mu}\hat{\partial}_\mu \Omega-18\hat{\partial}^{\mu}\Omega \hat{\partial}_\mu \Omega+6\Omega \hat{\Box}\Omega. 
\end{align}
As a consistency check, we also note that the combination $-T+B$ transforms as
\begin{align}
-T+B=\Omega^2 (-\hat{T}+\hat{B})-12\hat{\partial}^{\mu}\Omega \hat{\partial}_\mu \Omega+6\Omega \hat{\Box}\Omega
\end{align}
which using the relation~(\ref{RTB}), gives the correct transformation law for the Ricci scalar
\begin{align}
R=\Omega^2 \hat{R}-12\hat{\partial}^{\mu}\Omega \hat{\partial}_\mu \Omega+6\Omega \hat{\Box}\Omega.
\end{align}

As an aside, we note that in~\cite{Bamba} it was observed that the tensor defined as
\begin{align}
C^{\rho}{}_{\mu\nu}=T^{\rho}{}_{\mu\nu}+S^{\rho}{}_{\mu\nu} 
\end{align}
is conformally invariant, that is under a conformal transformation
\begin{align}
\hat{C}^{\rho}{}_{\mu\nu}=C^{\rho}{}_{\mu\nu}.
\end{align}
This can henceforth be thought of as the teleparallel equivalent of the Weyl tensor of general relativity. This tensor can also be used to construct a teleparallel version of conformal gravity.

\section{$f(T)$ gravity}

We will now review what happens when one attempts to conformally transform $f(T)$ gravity to a scalar frame. To do this we follow the approach used in~\cite{Yang:2010ji,Li:2010cg}. Let us start with the following $f(T)$ action
\begin{align}
S_{f(T)}= -\frac{ 1}{16\pi G}\int   f(T) e\, d^4x \,. \label{f(T)aux1}
\end{align}
Following the same procedure as the case of $f(R)$ gravity, we can introduce two auxiliary fields $\chi$ and $\phi$ such that the $f(T)$ action takes the following form
\begin{align}
S= -\frac{ 1}{16\pi G}\int   \left[\chi(T-\phi)+f(\phi) \right]e\, d^4x \,. \label{f(T)aux}
\end{align}
Varying this action with respect to $\chi$ yields $\phi=T$ showing that the action is indeed equivalent to the $f(T)$ action, unless $f''(T)\equiv 0$, in which case we are already working in Einstein gravity. Instead varying with respect to $\phi$ then yields $\chi=f'(\phi)$.

Now introducing the notation $F(\phi)=f'(\phi)$ , we can recast the theory into the following scalar-tensor type theory
\begin{align}
S= \frac{ 1}{16\pi G} \int  \left[-F(\phi) T- \omega(\phi)g^{\mu\nu}\nabla_\mu \phi \nabla_\nu \phi-V(\phi) \right]e\, d^4x \,. \label{tpbrans}
\end{align} 
In this particular case the kinetic term coefficient $\omega(\phi)$ is identically zero, $\omega(\phi)=0$, and the scalar field potential $V$ is given by
\begin{align}
V(\phi)=f(\phi)-\phi f'(\phi).
\end{align}

Now let us apply a general conformal transformation as outlined above~(\ref{conformal}). Then the action~(\ref{tpbrans}) transforms to the following
\begin{align}
S= \frac{ 1}{16\pi G} \int   \left[-F(\phi) (\Omega^{-2} \hat{T}-4\Omega^{-3} \hat{g}^{\mu\nu}\partial_\mu \Omega \hat{T}_\nu-6 \Omega^{-4}\hat{g}^{\mu\nu}\partial_\mu \Omega \partial_\nu \Omega)-\Omega^{-4}V(\phi) \right]\hat{e}\, d^4x \,. \label{tpconformal}
\end{align}
In order for the coupling between the gravitational sector and the scalar field to be minimal, we need to chose the conformal factor to be
\begin{align}
\Omega^2=F(\phi).
\end{align}
This means the action becomes
\begin{align}
S=  \frac{ 1}{16\pi G} \int \left[ ( -\hat{T}+2F(\phi)^{-1} \hat{g}^{\mu\nu}\partial_\mu F(\phi) \hat{T}_\nu+ \frac{3F'(\phi)^2}{2F(\phi)^2}\hat{g}^{\mu\nu}\partial_\mu \phi \partial_\nu \phi)-F^{-2}(\phi)V(\phi) \right]\hat{e}\, d^4x \,. \label{tpconformal0}
\end{align}

Now in order to get the kinetic term to be of the correct form, we define a new scalar field implicitly by
\begin{align}
\frac{d\psi}{d\phi}=\sqrt{3}\frac{F'(\phi)}{F(\phi)}, \label{psi}
\end{align}
which can be solved for $\psi$ to give
\begin{align}
\psi=\sqrt{3}\ln F(\phi).
\end{align}
This results in the action takes the following form
\begin{align}
S= \frac{ 1}{16\pi G}\int   \left[-\hat{T}+2F^{-1}\hat{\partial}_\mu F \hat{T}^\mu +\frac{1}{2}g^{\mu\nu}\nabla_\mu \psi \nabla_\nu \psi-U(\psi) \right]\hat{e}\, d^4x \, , \label{tpconformal1}
\end{align}
where the new potential $U$ is given by $U(\psi)=V(\phi)/F^2(\phi)$. 
We note that we have corrected a few algebraic errors which were presented in the original work of~\cite{Yang:2010ji}.

Now let us examine the second term in the action~(\ref{tpconformal}). Using that
\begin{align}
F^{-1}\hat{\partial}_\mu F=\hat{\partial}_\mu (\ln F),
\end{align}
we can integrate this term by parts to find that the action takes the following form
\begin{align}
S= \frac{ 1}{16\pi G}\int   \left[-\hat{T}-\ln(F)\hat{ B} +\frac{1}{2}g^{\mu\nu}\nabla_\mu \psi \nabla_\nu \psi-U(\psi) \right]\hat{e}\, d^4x \,. \label{tpconformal2}
\end{align}
Now the factor in front of $\hat{B}$ in the action can be expressed in terms of $\psi$, 
and so finally the action becomes
\begin{align}
S= \frac{ 1}{16\pi G}\int  \left[-\hat{T}-\frac{\psi}{\sqrt{3}}\hat{ B} +\frac{1}{2}g^{\mu\nu}\nabla_\mu \psi \nabla_\nu \psi-U(\psi) \right]\hat{e}\, d^4x \,. \label{tpconformal3}
\end{align}
this action represents a scalar field with a linear nonminimal coupling to a boundary term. However, as has been noted in~\cite{Li:2010cg} the kinetic energy term has the incorrect sign, thus the action is that of a phantom field, which generically leads to instabilities at the level of perturbations. 

This particular form of coupling is identical to the one first introduced in~\cite{Saez}. A very similar model was recently studied in~\cite{Bahamonde:2015hza}, where a quadratic coupling between $B$ and the scalar field, $\psi^2 B$, was present. Such a model exhibits interesting cosmological phenomenology, however, there the kinetic term was of the canonical form, so had a different sign to the above conformally transformed $f(T)$ action~(\ref{tpconformal3}), and thus was not subject to the instabilities that are present in $f(T)$ gravity. 


\section{Conformal equivalence of teleparallel dark energy and $f(T,B)$ gravity}
In this section we will work the other way around, that is we will start with a teleparallel scalar tensor model, and transform it into a modified gravity theory.

Let us begin with the following action, where a nonminimal coupling between a scalar field and the torsion scalar is present, sometimes referred to as {\it teleparallel dark energy}
\begin{align}
S= \frac{ 1}{16\pi G}\int  \left[ -A(\phi)T-\frac{1}{2}\partial_\mu \phi \partial^\mu \phi-V(\phi)+\mathcal{L}_m \right] e d^4x .
\end{align}
This action was first introduced in~\cite{Geng:2011aj} for the particular choice $A(\phi)=1+\xi \phi^2$, and a dynamical systems analysis of the model was performed in~\cite{Xu:2012jf,Wei:2011yr}. This was later generalised to include a more general coupling between $\phi$ and $T$ in~\cite{Otalora:2013dsa,Otalora:2013tba,Skugoreva:2014ena}. Immediately from the results of the previous section, we know that such a theory cannot be conformally transformed to an $f(T)$ gravity theory as there is no coupling between $\phi$ and $B$ present. However we will show in this section that it can be conformally transformed to the broader class of theories known as $f(T,B)$ gravity. 

Let us apply a conformal transformation to this theory in an attempt to remove the kinetic term from this action. A general conformal transformation changes the action to the following
\begin{align}
S=\frac{ 1}{16\pi G} \int \Omega^{-4} \left[- A(\phi)(\Omega^2 \hat{T}-4\Omega \partial_\mu \Omega \hat{T}^\mu-6 \hat{g}^{\mu\nu}\partial_\mu \Omega \partial_\nu \Omega)-\frac{1}{2}\Omega^2\hat{g}^{\mu\nu}\partial_\mu \phi \partial_\nu \phi-V(\phi) \right]  \hat{e}\, d^4x.
\end{align}
Now requiring that the kinetic term of the scalar field vanishes gives the following condition
\begin{align}
A(\phi)\left(\frac{d\Omega}{d\phi}\right)^2=\frac{1}{12}\Omega^2. \label{Adiffeqn}
\end{align}
Solving this will enable us to choose the conformal factor in terms of $\phi$
\begin{align}
\Omega=\exp\left({\int\frac{1}{2\sqrt{3A(\phi)}} d\phi}\right). \label{Omega}
\end{align}
We can also formally invert this relation meaning we can write $\phi$ as a function of $\Omega$, $\phi=\phi(\Omega)$. So now the action becomes
\begin{align}
S= \frac{ 1}{16\pi G}\int  \left[  -A(\Omega)\Omega^{-2}\hat{T}+4\Omega^{-3}A(\Omega) \partial_\mu \Omega \hat{T}^\mu-U(\Omega)\right] \hat{e}\, d^4x , \label{tpdark1}
\end{align}
where again the new potential $U(\Omega)$ is given simply by
\begin{align}
U(\Omega)=\frac{V(\phi)}{\Omega^4}.
\end{align}

It appears at this stage that the presence of $A(\Omega)$ in the second term of~(\ref{tpdark1}) ruins the possibility of this being equivalent to an $f(T,B)$ gravity. However if we introduce the function
\begin{align}
G(\Omega)= \int \frac{A(\Omega)}{\Omega^3} d\Omega,
\end{align}
we can write the second term of~(\ref{tpdark1}) as
\begin{align}
\Omega^{-3}A(\Omega) \partial_\mu \Omega=\partial_\mu G(\Omega).
\end{align}
Now we can integrate this term by parts so that the action takes the form
\begin{align}
S= \frac{ 1}{16\pi G}\int  \left[ - A(\Omega)\Omega^{-2}\hat{T}-2 G(\Omega) \hat{B}-U(\Omega) \right] \hat{e}\, d^4x.
\end{align}

Now the scalar field $\Omega$ has no kinetic term and is just an auxiliary field. Varying the action with respect to $\Omega$ and finding its equation of motion gives
\begin{align}
\frac{2A(\Omega)-\Omega A'(\Omega)}{\Omega^3}\hat{T}-\frac{2A(\Omega)}{\Omega^3}\hat{B}-U'(\Omega)=0. \label{omegatbeqn}
\end{align}
Now this can be formally solved to find $\Omega$ in terms of $\hat{T}$ and $\hat{B}$, $\Omega=\Omega(\hat{T},\hat{B})$ and so the action can be written as an $f(T,B)$ theory, with the function $f$ given by
\begin{align}
f(T,B)= -A(\Omega)\Omega^{-2}\hat{T}-2 G(\Omega) \hat{B}-U(\Omega). \label{ftb}
\end{align}
Thus we have established that a teleparallel dark energy theory with an arbitrary coupling between $T$ and the scalar field can be written as a particular instance of $f(T,B)$ gravity. In the next section we will derive conditions on the functional form of $f$ for such a teleparallel dark energy model. 

We note there is one particular class of models for which the functional form of the $f(T,B)$ gravity will take the form
\begin{align}
f(T,B)=-\alpha T+f(B),
\end{align}
so that we have an Einstein gravity plus some additional $f(B)$ contribution. This is when the coefficient of $\hat{T}$ vanishes in~(\ref{omegatbeqn}), so that one can invert $U'(\Omega)$ to find $\Omega$ as a function of $\hat{B}$ only. This is when the coupling function $A$ takes the form
\begin{align}
A(\Omega)=\alpha \Omega^2,
\end{align}
where $\alpha$ is some positive constant. Finding this originally in terms of $\phi$, we derive that the coupling function is given by
\begin{align}
A(\phi)=\beta^2(1+\frac{\phi}{2\beta\sqrt{3\alpha}})^2, \label{A}
\end{align}
where $\beta$ is some arbitrary constant.

To conclude this section let us give an explicit toy example of a nonminimally coupled teleparallel dark energy theory and transform it to an $f(T,B)$ gravity. Let us suppose the coupling function $A(\phi)$ is given by the simple form
\begin{align}
A(\phi)=\frac{\phi^2}{\sqrt{3}}.
\end{align}
This means that we can write $\Omega$ in terms of $\phi$, using~(\ref{Omega}), simply as
\begin{align}
\Omega=\sqrt{\phi}. 
\end{align}
Now let us choose a simple quadratic potential so that it can easily be inverted
\begin{align}
U(\phi)=m^2\phi^2=m^2 \Omega^4,
\end{align}
where $m$ is a constant. Then solving~(\ref{omegatbeqn}) for $\Omega$ gives
\begin{align}
\Omega=\left(\frac{T+B}{2\sqrt{3} m^2}\right)^{1/2}.
\end{align}
Inserting this back into~(\ref{ftb}) will give us the following functional form of $f(T,B)$
\begin{align}
f(T,B)=\frac{1}{12m^2}(T+B)^2.
\end{align}

\section{$f(T,B)$ gravity to scalar tensor theory}

In this section we will explore the consequences when one conformally transforms $f(T,B)$ gravity to a scalar frame. We start with the gravitational sector of the $f(T,B)$ action
\begin{align}
  S_{\rm f(T,B)} = \frac{ 1}{16\pi G}\int 
   f(T,B)
   e\, d^4x \,,\label{actionftb2}
\end{align}
and we introduce the four auxiliary fields $\chi_1$, $\chi_2$, $\phi_1$ and $\phi_2$, writing the above action in the equivalent form
\begin{align}
 S =\frac{ 1}{16\pi G} \int  
    \left[f(\phi_1,\phi_2)+ \chi_1(T-\phi_1)+ \chi_2(B-\phi_2)
  \right] e\, d^4x \,.\label{actionftbaux}
\end{align}
Varying with respect to $\chi_1$ yields $T=\phi_1$ and with respect to $\chi_1$ gives $B=\phi_2$. Now varying with respect to $\phi_1$ and $\phi_2$ give $\chi_1=f^{(1,0)}(\phi_1,\phi_2)$ and $\chi_2=f^{(0,1)}$ respectively. This leaves the following action
\begin{align}
 S =\frac{ 1}{16\pi G} \int  
    \left[f(\phi_1,\phi_2)+ (T-\phi_1)f^{(1,0)}(\phi_1,\phi_2)+ (B-\phi_2)f^{(0,1)}(\phi_1,\phi_2)
  \right] e\, d^4x \,,\label{actionftbaux0}
\end{align}
assuming that neither of the second derivatives $f^{(2,0)}$ or $f^{(0,2)}$ vanishes (the case of $f^{(0,2)}=0$ is equivalent to $f(T)$ gravity and was examined in Section IV and we will cover the remaining case in the next section). We can rewrite this slightly differently as the following scalar tensor type action with two scalar fields
\begin{align}
S =\frac{ 1}{16\pi G} \int  
    \left[- F(\phi_1,\phi_2)T + G(\phi_1,\phi_2)B-V(\phi_1,\phi_2)
  \right] e\, d^4x \,,\label{actionftbaux1}
\end{align}
where we introduce the notation $F(\phi_1,\phi_2)=-f^{(1,0)}(\phi_1,\phi_2)$ and $G(\phi_1,\phi_2)=f^{(0,1)}(\phi_1,\phi_2)$. The double potential $V(\phi_1,\phi_2)$ is given by
\begin{align}
V(\phi_1,\phi_2)=\phi_1 f^{(1,0)}(\phi_1,\phi_2)+\phi_2f^{(0,1)}(\phi_1,\phi_2)-f(\phi_1,\phi_2).
\end{align}
This is a particular instance of a teleparallel scalar-tensor theory with two scalar fields. Conformal transformations with multiple scalar fields have been discussed in~\cite{Kaiser:2010ps,Abedi:2014mka}. 

Now let us apply a conformal transformation to the action~(\ref{actionftbaux1}). We find
\begin{multline}
 S = \frac{1}{16\pi G}\int\,\left[\right.
    (-\Omega^{-2} \hat{T}+4\Omega^{-3} \hat{\partial}_\mu \Omega \hat{T}^\mu+6 \Omega^{-4}\hat{\partial}_\mu \Omega \hat{\partial}^\nu \Omega)F(\phi_1,\phi_2)\\+( \Omega^{-2} \hat{B}-4\Omega^{-3} \hat{T^\mu}\hat{\partial}_\mu \Omega+6\Omega^{-3} \hat{\partial}^\mu \hat{\partial}_\mu\Omega-18\Omega^{-4}\hat{\partial}^{\mu}\Omega \hat{\partial}_\mu \Omega+\frac{6}{\hat{e}}\Omega^{-3} \hat{\partial}^\mu \Omega \hat{\partial}_\mu \hat{e})G(\phi_1,\phi_2)
  -\Omega^{-4}V(\phi_1,\phi_2)\left.\right] \hat{e} \, d^4x\,,\label{actionftbaux2}
\end{multline}
We can integrate the term with $\partial_\mu \hat{e}$ by parts to obtain
\begin{multline}
 S = \frac{1}{16\pi G}\int\, (
    (-\Omega^{-2} \hat{T}+4\Omega^{-3} \hat{\partial}_\mu \Omega \hat{T}^\mu+6 \Omega^{-4}\hat{\partial}_\mu \Omega \hat{\partial}^\nu \Omega)F(\phi_1,\phi_2)\\+ ( \Omega^{-2} \hat{B}-4\Omega^{-3} \hat{T^\mu}\hat{\partial}_\mu \Omega )G(\phi_1,\phi_2)-6\Omega^{-3}\hat{\partial}^\mu \Omega \partial_\mu G(\phi_1,\phi_2)
  -U(\phi_1,\phi_2)) \hat{e}\, d^4x \,,\label{actionftbaux3}
\end{multline}
where we have disregarded a boundary term and we also introduce the new potential $U(\phi_1,\phi_2)=\Omega^{-4}V(\phi_1,\phi_2)$. 

We want to explore under what conditions one can choose a suitable conformal factor to eliminate the coupling between either $T$ or the boundary term $B$. It is straightforward to observe that it is always possible to eliminate the coupling between the scalar field and the torsion scalar $T$, one simply chooses the conformal factor to be $\Omega^2=F(\phi_1,\phi_2)$. However, to eliminate couplings between the scalar fields and $B$, or equivalently the vector $T^\mu$, requires a longer calculation. Integrating the boundary term by parts, we get left with the following coefficient of the vector $T^\mu$ in the above action~(\ref{actionftbaux3})
\begin{align}
\left(4\Omega^{-3} \hat{\partial}_\mu \Omega (F(\phi_1,\phi_2)-G(\phi_1,\phi_2))-2\partial_\mu(\Omega^{-2}G(\phi_1,\phi_2))\right)T^\mu
\nonumber\\=\left(4\Omega^{-3} \hat{\partial}_\mu \Omega F(\phi_1,\phi_2)-2\Omega^{-2}\partial_\mu(G(\phi_1,\phi_2))\right)T^\mu.
\end{align}
And thus a sufficient condition for the coupling between $T^\mu$ to vanish is that
\begin{align}
2\Omega^{-1} \hat{\partial}_\mu \Omega F(\phi_1,\phi_2)-\partial_\mu(G(\phi_1,\phi_2))=0. \label{omegaeqn}
\end{align}
Now can we choose a sufficient $\Omega=\Omega(\phi_1,\phi_2)$ such that this will vanish? Expanding~(\ref{omegaeqn}) in terms of $\phi_1$ and $\phi_2$ partial derivatives gives the following two first order partial differential equations
\begin{align}
2\Omega^{-1}  \Omega^{(1,0)} F(\phi_1,\phi_2)-G^{(1,0)}(\phi_1,\phi_2)&=0\\
2\Omega^{-1}  \Omega^{(0,1)} F(\phi_1,\phi_2)-G^{(0,1)}(\phi_1,\phi_2)&=0
\end{align}
which we can rewrite as
\begin{align}
\Omega^{(1,0)}&=\frac{\Omega}{2F}G^{(1,0)} \label{pde1}
\\
\Omega^{(0,1)}&=\frac{\Omega}{2F}G^{(0,1)}. \label{pde2}
\end{align}
Now for such a solution $\Omega$ to exist for these partial differential equations, we simply require that the second mixed derivatives agree, that is if we differentiate the first of these equations with respect to $\phi_2$ it must equal the second equation differentiated with respect to $\phi_1$. After doing this calculation, we find the following condition on our original function $f$, which must be satisfied in order for such an $\Omega$ to exist
\begin{align}
f^{(2,0)}f^{(0,2)}=(f^{(1,1)})^2.
\end{align}
One such solution to this equation is $f(R)$ gravity, when $f^{(1,0)}=-f^{(0,1)}$, but the equation has other solutions too, including separable solutions. 

Finally we mention, for the couplings between both $T$ and $T^\mu$ to simultaneously vanish, we require $\Omega=F^{1/2}$ and the system~(\ref{pde1})-(\ref{pde2}) to hold. But then solving for these two conditions requires that $F=-G$, which is simply the case of the teleparallel equivalent of $f(R)$ gravity, when the functional form of $f(T,B)$ is $f(T,B)=f(-T+B)=f(R)$.  And so the unique class of $f(T,B)$ gravity which has an Einstein frame is $f(R)$ gravity, as to be expected. 

\section{$f(B)$ gravity}

For completeness, in this section we will examine the final case of $f(T,B)$ gravity we have yet to examine. This is the case when we have the following action
\begin{align}
  S_{\rm f(T,B)} = \frac{ 1}{16\pi G}\int \left[
   \alpha T +f(B) \right]
   e\, d^4x \,,\label{actionfb}
\end{align}
and without loss of generality we will set $\alpha=-1$ so that the action takes the form of Einstein gravity plus a boundary term modification. This action was not covered by the analysis in the previous section since for this particular action $f_{TT}=0$. Performing the standard transformation using auxiliary variables, the action can be recast into the following form
\begin{align}
  S_{\rm f(T,B)} = \frac{ 1}{16\pi G}\int 
   \left[-T +F(\phi)B- V(\phi)
   \right]e\, d^4x \,,\label{actionfb1}
\end{align}
where $F(\phi)=f'(\phi)$ and $V(\phi)=\phi f'(\phi)-f(\phi)$. 

Let us attempt to remove the coupling between $\phi$ and $B$ in this action. Now applying a conformal transformation and performing some integration by parts recasts this into the following form
\begin{align}
S = \frac{1}{16\pi G}\int\,\left[
    -\Omega^{-2} \hat{T}+\Omega^{-2}(1+G(\phi))\hat{B}-4\Omega^{-3} \hat{T^\mu}\hat{\partial}_\mu \Omega F(\phi)+6 \Omega^{-4}\hat{\partial}_\mu \Omega \hat{\partial}^\nu \Omega-6\Omega^{-3}\hat{\partial}^\mu \Omega \hat{\partial}_\mu F(\phi)
  -U(\phi)\right]     \hat{e}\,d^4 x \,,
\end{align}
where the new potential $U(\phi)=\Omega^{-4} V(\phi)$. We now explore whether or not we can choose a suitable $\Omega=\Omega(\phi)$ to remove the couplings between $\phi$ and both $\hat{B}$ and $\hat{T}^\mu$. Rewriting the partial derivatives in terms of $\phi$ and integrating by parts the term with $\hat{T}^\mu$ gives
\begin{align}
S = \frac{1}{16\pi G}\int\,\left[
    -\Omega^{-2} \hat{T}+\Omega^{-2}(1+F(\phi)+\Omega^2H(\phi))\hat{B}+\left(6 \Omega^{-4}\left(\frac{d\Omega}{d\phi}\right)^2-6\Omega^{-3}F'(\phi)\frac{d\Omega}{d\phi}\right)\hat{\partial}^\mu \phi \hat{\partial}_\mu \phi
  -U(\phi)\right]     \hat{e}\,d^4 x \,, \label{fb2}
\end{align}
where $H(\phi)$ is given by the following integral
\begin{align}
H(\phi)=2\int \frac{F(\phi)}{\Omega(\phi)^3}\frac{d\Omega}{d\phi} d\phi.
\end{align}

In order for the boundary term to have no effect on the action, we require that the coefficient of $\hat{B}$ is simply a constant $\beta$, so that
\begin{align}
\Omega^{-2}(1+F(\phi)+\Omega^2H(\phi))=\beta,
\end{align}
This will allow us to find $\Omega$ in terms of $\phi$. Differentiating this with respect to $\phi$ gives
\begin{align}
-\frac{2}{\Omega^3}\frac{d\Omega}{d\phi}+\frac{F'(\phi)}{\Omega^2}=0.
\end{align}
And so solving this for $\Omega$ gives
\begin{align}
\Omega= e^{F(\phi)/2}.
\end{align}

Inserting this solution for $\Omega$  into our action~(\ref{fb2}) gives us
\begin{align}
S = \frac{1}{16\pi G}\int\,\left[
    -e^{-F(\phi)} \hat{T}-\frac{3}{2}\left( e^{-F(\phi)}F'(\phi)^2\right)\hat{\partial}^\mu \phi \hat{\partial}_\mu \phi
  -U(\phi)\right]     \hat{e}\,d^4 x \,. \label{fb3}
\end{align}
Finally, introducing the new scalar field $\varphi= 2\sqrt{3}(e^{-F(\phi)/2}-\beta)$ recasts the action into the following form
\begin{align}
S = \frac{1}{16\pi G}\int\,\left[
    - A(\varphi)\hat{T}-\frac{1}{2}\hat{\partial}^\mu \varphi \hat{\partial}_\mu \varphi
  -U(\varphi)\right]     \hat{e}\,d^4 x \,, \label{fb4}
\end{align}
where the coupling function $A(\varphi)$ is given by
\begin{align}
A(\varphi)=\beta^2(1+\frac{\varphi}{2\sqrt{3}\beta})^2,
\end{align}
in agreement with the result~(\ref{A}) found in Section IV. Thus we have found that $f(B)$ gravity, where we have an additional $f(B)$ term added to the Einstein Hilbert action, is conformally equivalent to a particular instance of teleparallel dark energy. Moreover, as opposed to $f(T)$ gravity, the kinetic term has the correct sign and so will not suffer from potential instabilities to perturbations.  

\section{Discussion}

\begin{figure}[!b]
\begin{tikzpicture}
  \matrix (m) [matrix of math nodes,row sep=3em,column sep=2.5em,minimum width=1em]
  {& & f(T,B) &  & \\
   \textrm{Functional form:} & f(B) & f(T) &  f_{TT}f_{BB}=f_{TB}^2 & \\
   \textrm{Coupling after } g\rightarrow \Omega^2 g :  & \beta^2(1+\frac{\varphi}{2\sqrt{3}\beta})^2 T & -\frac{\phi}{\sqrt{3}} B  & A(\phi) T & \textrm{No coupling} \\};
  \path[-stealth]
    (m-1-3) edge [dashed,-]  (m-2-2)
    (m-1-3) edge [dashed,-] (m-2-3)
    (m-1-3) edge [dashed,-] (m-2-4)
    (m-2-2) edge node [below] {} (m-3-2)
    (m-2-3) edge node [right] {} (m-3-3)  
    (m-2-4) edge node [right] {} (m-3-4)
    (m-2-4) edge node [right] {\, $f(T,B)=f(-T+B)$} (m-3-5)    ;
\end{tikzpicture}
\caption{The conformal equivalence of different $f(T,B)$ gravity models. The top line shows the particular functional form of the $f(T,B)$ gravity considered, and the bottom line shows the type of nonminimal coupling present in the action after a conformal transformation. A minimal coupling is only possible in the case when $f(T,B)=f(R)$. The kinetic and potential energy of the scalar field are also present in the conformally transformed action, with the kinetic term being either the canonical or phantom type. }
\label{pic}
\end{figure}
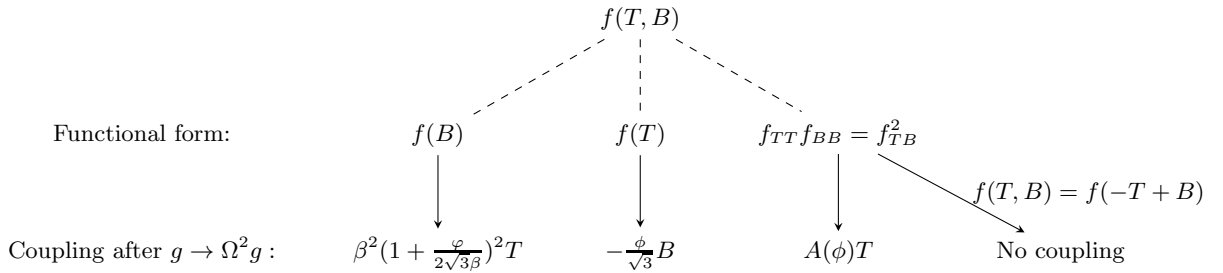

In this work we have explored the conformal relationships between various modified teleparallel gravity theories. Conformally transforming these theories indicates the need to take into account first derivatives of the torsion, in particular the scalar $B$ given by the divergence of a contraction of the torsion tensor is needed to be taken into account. We first reviewed $f(T)$ gravity, showing that it is conformally equivalent to a phantom teleparallel scalar tensor theory with a linear nonminimal coupling between the scalar field and the boundary term only, similar to a model recently proposed in~\cite{Bahamonde:2015hza}. 

Furthermore we looked at a teleparallel dark energy theory, where a generic nonminimal coupling between the scalar field and the torsion scalar $T$ was present. We showed that in general this is conformally equivalent to a particular $f(T,B)$ gravity, moreover if this $f(T,B)$ gravity is nonlinear in both $T$ and $B$ then it must satisfy the condition
\begin{align}
f_{TT}f_{BB}&=(f_{TB})^2. \label{condition}
\end{align}
We gave an explicit toy example of such a coupling and transformed it into a simple $f(T,B)$ theory, which did indeed satisfy condition~(\ref{condition}). The other possibility is that the coupling between $\phi$ and $T$ takes the particular form
\begin{align}
A(\phi)=\beta^2(1+\frac{\phi}{2\beta\sqrt{3\alpha}})^2,
\end{align}
in which case the model is conformally equivalent to a particular $f(T,B)$ theory of the form
\begin{align}
f(T,B)=-\alpha T+f(B),
\end{align}
with only a linear dependence on the torsion scalar, and the particular $f(B)$ is dependent on the structure of the of the potential of the scalar field.  

Moreover, we also looked at the case of $f(B)$ gravity, where the functional form of $f(T,B)$ gravity is given by $f(T,B)=-T+f(B)$. It was shown that this can always be conformally transformed to a frame where there is a particular type of nonminimal coupling between the scalar field and $T$. Thus we have an interesting duality relation, $f(T)$ can always be transformed to a $A(\phi)B$ phantom scalar field theory, whereas $f(B)$ can always be transformed to a canonical $A(\phi)T$ theory.  

We have derived various relationships between modified teleparallel theories of gravity and teleparallel scalar-tensor theories. The unique form of these different theories which has an Einstein frame is given by $f(T,B)$ gravity which takes the form $f(-T+B)$, which is equivalent to $f(R)$~\cite{Bahamonde:2015zma}. In all other cases a form of nonminimal coupling between the scalar field and the gravitational sector remains present. The various conformal relationships between the different theories considered have been summarised in Figure~\ref{pic}.

\begin{acknowledgments}
The author would like to thank Christian B\"ohmer for invaluable feedback and Sebastian Bahamonde for useful discussions.  
\end{acknowledgments}


\begin{thebibliography}{99}
\bibitem{Briscese:2006xu}
  F.~Briscese, E.~Elizalde, S.~Nojiri and S.~D.~Odintsov,
  Phys.\ Lett.\ B {\bf 646} (2007) 105
  [hep-th/0612220].


\bibitem{Ferraro:2006jd}
  R.~Ferraro and F.~Fiorini,
  Phys.\ Rev.\ D {\bf 75} (2007) 084031
  [gr-qc/0610067].

\bibitem{Bengochea:2008gz}
  G.~R.~Bengochea and R.~Ferraro,
  Phys.\ Rev.\ D {\bf 79} (2009) 124019
  [arXiv:0812.1205 [astro-ph]].

\bibitem{Li:2010cg}
  B.~Li, T.~P.~Sotiriou and J.~D.~Barrow,
  Phys.\ Rev.\ D {\bf 83} (2011) 064035
  [arXiv:1010.1041 [gr-qc]].
   
\bibitem{Krssak:2015lba}
  M.~Kr\v{s}\v{s}\`ak,
  arXiv:1510.06676 [gr-qc].

\bibitem{Krssak:2015oua}
 M.~Kr\v{s}\v{s}\`ak and E.~N.~Saridakis,
  arXiv:1510.08432 [gr-qc].
  
  
\bibitem{Geng:2011aj}
  C.~Q.~Geng, C.~C.~Lee, E.~N.~Saridakis and Y.~P.~Wu,
  Phys.\ Lett.\ B {\bf 704} (2011) 384
  [arXiv:1109.1092 [hep-th]].

\bibitem{Bahamonde:2015zma}
  S.~Bahamonde, C.~G.~B\"ohmer and M.~Wright,
  Phys.\ Rev.\ D {\bf 92} (2015) 10,  104042
  [arXiv:1508.05120 [gr-qc]].


 \bibitem{Bahamonde:2015hza}
  S.~Bahamonde and M.~Wright,
  Phys.\ Rev.\ D {\bf 92} (2015) 8,  084034
  [arXiv:1508.06580 [gr-qc]].
  
  
\bibitem{Zubair:2016uhx}
  M.~Zubair and S.~Bahamonde,
  arXiv:1604.02996 [gr-qc].
  
\bibitem{Saez}
  D.~Saez,
  Phys.\ Rev.\ D {\bf 27} (2015) 12,  2839  
  
\bibitem{Moller}
C.~Moller,
 Kong.~Dan.~Vid.~Sel.~Mat.~Fys.~Med. {\bf 89}, (1978) 13  
   

\bibitem{Yang:2010ji}
  R.~J.~Yang,
  Europhys.\ Lett.\  {\bf 93} (2011) 60001
  [arXiv:1010.1376 [gr-qc]].
    
  
\bibitem{Bamba}
K. Bamba, S. D. Odintsov and D. S\'{a}ez-G\'{o}mez,
Phys. Rev. D {\bf 88}, 084042 (2013)
[arXiv:1308.5789 [gr-qc]]

\bibitem{Otalora:2014aoa}
  G.~Otalora,
  Int.\ J.\ Mod.\ Phys.\ D {\bf 25} (2015) no.02,  1650025
  doi:10.1142/S0218271816500255
  [arXiv:1402.2256 [gr-qc]].

  
\bibitem{Faraoni:1998qx}
  V.~Faraoni, E.~Gunzig and P.~Nardone,
  Fund.\ Cosmic Phys.\  {\bf 20} (1999) 121
  [gr-qc/9811047].
  
  
\bibitem{Xu:2012jf}
  C.~Xu, E.~N.~Saridakis and G.~Leon,
  JCAP {\bf 1207} (2012) 005
  [arXiv:1202.3781 [gr-qc]].
  
  
\bibitem{Wei:2011yr}
  H.~Wei,
  Phys.\ Lett.\ B {\bf 712} (2012) 430
  [arXiv:1109.6107 [gr-qc]].
  
  
\bibitem{Otalora:2013tba}
  G.~Otalora,
  JCAP {\bf 1307} (2013) 044
  [arXiv:1305.0474 [gr-qc]].
  
  
\bibitem{Otalora:2013dsa}
  G.~Otalora,
  Phys.\ Rev.\ D {\bf 88} (2013) 063505
  [arXiv:1305.5896 [gr-qc]].
  
  
\bibitem{Skugoreva:2014ena}
  M.~A.~Skugoreva, E.~N.~Saridakis and A.~V.~Toporensky,
  Phys.\ Rev.\ D {\bf 91} (2015) 044023
  [arXiv:1412.1502 [gr-qc]].
  
\bibitem{Li:2011rn}
  M.~Li, R.~X.~Miao and Y.~G.~Miao,
  JHEP {\bf 1107} (2011) 108
  doi:10.1007/JHEP07(2011)108
  [arXiv:1105.5934 [hep-th]].
  
 \bibitem{Kaiser:2010ps}
  D.~I.~Kaiser,
  Phys.\ Rev.\ D {\bf 81} (2010) 084044
  [arXiv:1003.1159 [gr-qc]].
    
    
\bibitem{Abedi:2014mka}
  H.~Abedi and A.~M.~Abbassi,
  JCAP {\bf 1505} (2015) no.05,  026
  doi:10.1088/1475-7516/2015/05/026
  [arXiv:1411.4854 [gr-qc]].
  
  
\bibitem{Obukhov:1982zn}
  Y.~N.~Obukhov,
  Phys.\ Lett.\ A {\bf 90} (1982) 13.
  doi:10.1016/0375-9601(82)90037-8

\end{thebibliography}
\end{document}